\documentclass[letter]{aa}
\usepackage{graphicx}
\usepackage{txfonts}
\usepackage[dvipsnames]{xcolor}
\usepackage{amssymb}
\usepackage{siunitx}
\usepackage{soul}
\usepackage{ulem}
\usepackage[hidelinks]{hyperref}
\hypersetup{
    colorlinks=true,
    linkcolor=blue,
    filecolor=blue,      
    urlcolor=blue,
    citecolor=blue,
    }

\begin{document}
\title{Quiet, but not silent}
\subtitle{The X-ray activity of the Maunder minimum star HD\,166620}

\author{M. M. Bennedik\inst{1} \and B. Stelzer\inst{1} \and 
H. Isaacson\inst{2} \and A. Binks\inst{1} \and M. Caramazza\inst{1}  \and F. Haberl\inst{3}}

\institute{$^1$Institut für Astronomie und Astrophysik, Eberhard Karls Universität Tübingen, Sand 1, 72076 Tübingen, Germany\\
\email{bennedik@astro.uni-tuebingen.de}\\
  $^2$ 501 Campbell Hall, University of California at Berkeley, Berkeley, CA 94720, USA\\
  $^3$ Max-Planck-Institut für extraterrestrische Physik, Giessenbachstrasse 1, 85748 Garching, Germany}
  
\date{Received ; accepted }

  \abstract{As the only known unambiguous star in a Maunder minimum-like chromospheric activity state, the properties of HD\,166620 can provide valuable insight into the behaviour of the Sun during the historic extended low-states of its activity cycle.
  The coronal X-ray activity of HD\,166620 has so far only been probed with a ROSAT/HRI observation in 1996, near the chromospheric activity maximum before the star entered its grand minimum around 2004. We conducted a deep {\it XMM-Newton} observation of HD\,166620 during its chromospheric Ca\,II~H\&K activity grand minimum to achieve a better understanding of its magnetic activity. 
  We detected HD\,166620 with an X-ray luminosity of ${{\rm log}\,L_{\rm X}\,\rm{(erg\,s^{-1})}=26.56^{+0.10}_{-0.12}}$, corresponding to ${{\rm log}\,(L_{\rm X}/L_{\rm bol}) = -6.58^{+0.10}_{-0.12}}$ and an X-ray surface flux of ${{\rm log}\,F_{\!\rm X}\,\rm{(erg\,cm^{-2}\,s^{-1})} = 3.97^{+0.10}_{-0.12}}$. With respect to the earlier ROSAT observation, the X-ray brightness of HD\,166620 has decreased by a factor of 2.5 during its Maunder minimum-like state. To place its X-ray properties into context, we constructed an X-ray sample of late-type stars within 10\,pc of the Sun.
  The activity of HD\,166620 is below the levels of all other K dwarfs in the 10\,pc sample. The corona of HD\,166620 during its grand minimum emits at the level of the solar background corona, which implies that it has no large active magnetic structures. Along with long-term Ca\,II~H\&K monitoring of HD\,166620, this result provides evidence that the solar activity during the Maunder minimum was not reduced significantly below the levels seen during its present-day cycle minima. The similar X-ray surface flux of HD\,166620 and the modern quiet Sun, and also their Rossby number near the critical value of spin-down models, suggest a connection between the regime of weakened magnetic braking and the occurrence of Maunder minimum states.}

   \maketitle

\section{Introduction}\label{sect:introduction}
During the solar Maunder minimum (henceforth MM) between 1645--1715, historical records show an extended period with significantly decreased sunspot counts and auroral sightings \citep{Usoskin_2015}. The cause for this extended low-activity state is still unknown and might have profound implications for our understanding of the underlying mechanisms of stellar dynamos and its effects on the climates of Earth and exoplanets. Furthermore, it is still to be determined whether the solar magnetic field during the MM was extremely weak or if it was near the cycle-minimum amplitude \citep{Baum_2022}.

In order to constrain stellar properties during MM states, it is crucial to confirm their occurrence in other Sun-like stars. The systematic search for extrasolar activity cycles began with the Mount Wilson HK survey \citep{Baliunas_1995}, in which chromospheric Ca\,II~H\&K lines of 111 stars with spectral types (SpT) F to M were monitored for three decades. The ratio of the fluxes of the H- and K-emission lines and the continuum on either side defines the $S_{\!{\rm HK}}$-index, which is widely used to characterise chromospheric activity \citep{Vaughan_1978}. The Mount Wilson survey revealed a high number of "flat" stars with a low and constant $S_{\!{\rm HK}}$-index \citep{Baliunas_1995}.
Initially flagged as potential MM-like states, most of these targets were instead found in their post-main-sequence subgiant phase, which suggests that MM states might be much rarer than was assumed before \citep{Wright_2004}.

To date, only a few MM candidates have been identified. Due to declining magnetic activity, HD\,4915 was proposed as an MM candidate by \citet{Shah_2018}, but this was refuted by \citet{Flores_Trivigno_2024} based on further monitoring. Other reported candidates are HD\,217014 \citep{Poppenhaeger_2009}, HD\,20807 \citep{Flores_2021}, and TIC\,352227373 \& TYC\,7560-477-1 \citep{Jaervinen_2025}. Whether these stars are in an MM or are generally inactive remains inconclusive, however. The only known star that unambiguously is in an MM-like state is HD\,166620. Five decades of data from long-term $S_{\!{\rm HK}}$-index monitoring clearly show that its pronounced activity cycle stopped \citep{Baum_2022, Luhn_2022}.

We provide the first X-ray detection of the K2 dwarf HD\,166620 since the start of its reduced chromospheric activity approximately 20 years ago. For this purpose, we conducted a dedicated observation with 
\textit{XMM-Newton}. We determine the X-ray surface flux of HD\,166620 and compare it with the surface fluxes of different types of magnetic structures seen on the Sun. To place the observed activity level of HD\,166620 into context, we construct an X-ray sample of all FGK dwarfs within 10\,pc based on the work of \citet{Reyle_2021}. In Sect.~\ref{sect:data} we describe the $S_{\!{\rm HK}}$ and X-ray data and analysis for HD\,166620 and the 10\,pc sample. We present and discuss the results of our analysis in Sect.~\ref{sect:discussion}, and we summarise our findings in Sect.~\ref{sect:summary}.
\begin{figure*}
    \centering
    \includegraphics[width=0.85\textwidth]{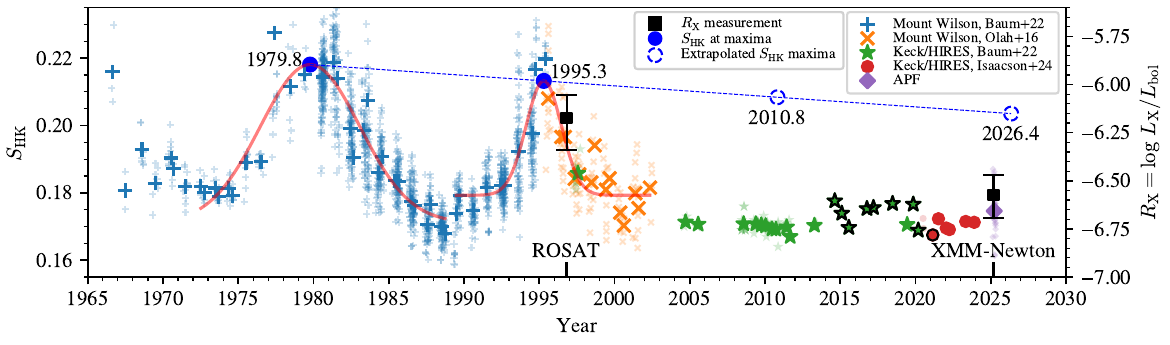}
    \caption{$S_{\!{\rm HK}}$ and $R_{\rm X}$ time series for HD\,166620 with data compiled by \citet{Luhn_2022} and more recent data from Keck/HIRES \citep{Isaacson_2024} and the APF \citep{Vogt_2014}. The small transparent symbols mark  individual $S_{\!{\rm HK}}$ measurements, and the larger opaque symbols are the same data, binned in 120\,d bins. A black outline denotes a bin containing only a single measurement. The dashed blue line shows the extrapolated trend of the two maxima (see Sect.~\ref{subsect:data-SHK}). The two X-ray measurements are overplotted with black squares, and their epochs are marked on the axis.
    }
    \label{fig:activity-timeseries}
\end{figure*}
\section{Data and analysis}\label{sect:data}
\subsection{$S_{\!{\rm HK}}$ activity}\label{subsect:data-SHK}
Long-term measurements of the $S_{\!{\rm HK}}$ activity of HD\,166620 have been discussed by \citet{Luhn_2022}, who used Mount Wilson and Keck/HIRES data from \citet{Baum_2022} and \citet{Olah_2016}. 
Since then, more Keck/HIRES data were published by \citet{Isaacson_2024}. For the time of our \textit{XMM-Newton} observation (which is described in Sect.~\ref{subsect:data-Xray}), we obtained data from the monitoring program of the Lick Automated Planet Finder (APF) \citep{Vogt_2014}. We discuss the APF data in Appendix~\ref{app:SHK-APF}.

The complete $S_{\!{\rm HK}}$-index time series of HD\,166620 is presented in Fig.~\ref{fig:activity-timeseries}. We binned the data in 120\,d steps and fitted the two observed cycle maxima with Gaussian functions. The Gaussians were not meant to represent a physical model of the variability and were only used to determine the time of the maxima. The peaks of the two Gaussians are separated by 15.5\,years, which agrees with the cycle period of $15.8 \pm 0.3$\,years found by \citet{Baliunas_1995}. Extrapolation predicts the subsequent maxima to occur in late-2010 and mid-2026,  potentially with a trend of decreasing peak $S_{\!{\rm HK}}$ values, as indicated by the dashed blue line in Fig.~\ref{fig:activity-timeseries}. No activity increase is visible in the $S_{\!{\rm HK}}$ data after the decrease from the 1995 maximum, however. Our new data from the APF still show no systematically increased activity, even though they were acquired only approximately one year before the predicted 2026 maximum. During the extended low state, the $S_{\!{\rm HK}}$ level is similar to the pre-MM cycle minima.

\subsection{X-ray activity}\label{subsect:data-Xray}
The only X-ray detection of HD\,166620 to date is from a ROSAT/HRI observation on 26~October~1996 \citep{Schmitt_2004}, shortly after the last maximum of its $S_{\!{\rm HK}}$ cycle. We conducted a dedicated \textit{XMM-Newton} observation on 10~March~2025 to quantify the X-ray activity level during its MM-like state.

\begin{figure}
    \centering
    \includegraphics[width=0.45\textwidth]{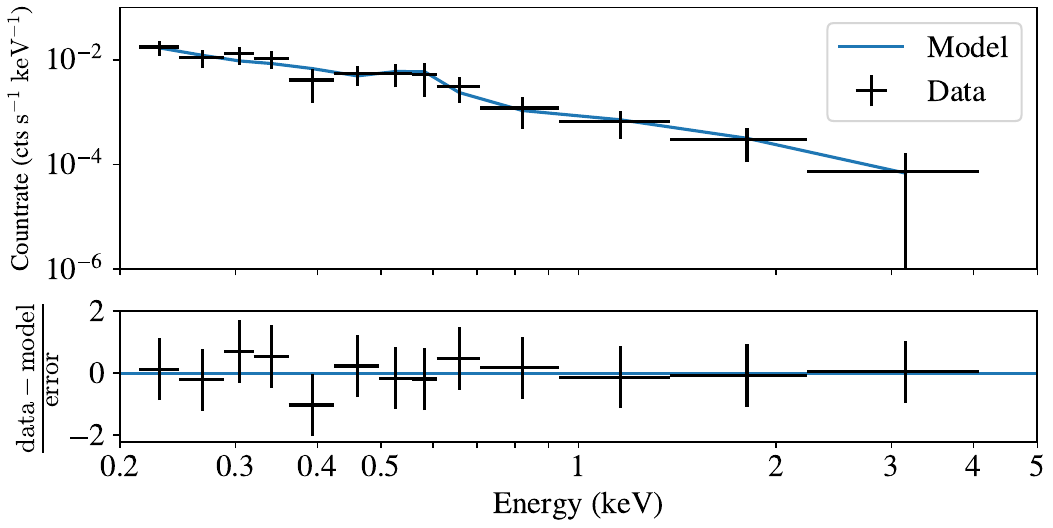}
    \caption{Top: \textit{XMM-Newton} EPIC/pn spectrum of HD\,166620 and best-fit 2T-APEC thermal model. Bottom: Residuals.}
    \label{fig:Maunder-Spec}
\end{figure}
For the data analysis, we used the \textit{XMM-Newton} Science Analysis Software (SAS) version 22.0.0. Given the low X-ray brightness of HD\,166620, we only considered the data from the most sensitive \textit{XMM-Newton} instrument, EPIC/pn. To eliminate effects of solar background irradiation, we filtered out time intervals with high-energy ($\geq$10\,keV) count rates on the whole detector above ${0.3\,\rm{ct/s}}$, which left a net total observation time of ${28.9\,\rm{ks}}$. Additionally, we excluded bad pixels (flag = 0) and retained events with pattern~$\leq$\,4. We ran the source detection, and the number of net source counts is ${87.4 \pm 12.9}$ for the full 0.2--12.0\,keV energy band of EPIC/pn.

We extracted a spectrum and the corresponding response files from a source region with a radius of $30^{\prime\prime}$ centred on the position of HD\,166620 at the epoch of the observation, which matches the position from the source detection within $0.1^{\prime\prime}$. The region we used for background-subtraction has a radius of $50^{\prime\prime}$ and is located on the same CCD. Then, we performed the spectral analysis with XSPEC\footnote{See: \url{https://heasarc.gsfc.nasa.gov/xanadu/xspec/}} version 12.14.1 using a spectral binning of $\geq$15 counts. We fitted a two-temperature thermal model \citep[APEC,][]{Smith_2001}. Using the library from \citet{Asplund_2009}, we set the global abundance of the stellar corona to a typical value of $Z=0.3Z_\sun$ \citep[see e.g.][]{Robrade_2005, Maggio_2007}, leaving the coronal plasma temperatures and the emission measures of the two components free to be fit. At the lower end, the energy range considered for the fit was limited to $\geq$0.2\,keV because the calibration of EPIC is unreliable at lower energies. At the upper end, we limited the spectrum to $\leq$5.0\,keV because the count rate of HD\,166620 is low at higher energies. The best-fit model yields a coronal plasma temperature of ${kT = 0.109^{+0.019}_{-0.015} \,\rm{keV}}$ for the cooler component. 
While the low number of counts prevented us from constraining temperature of the hotter component well, a statistically significant improvement over a one-temperature model is visible. The Fisher test \citep{Fisher_1922} yielded a p-value corresponding to a significance of $3.28\sigma$.
The spectrum and best-fit model are presented in Fig.~\ref{fig:Maunder-Spec}. The hotter component of the two-temperature model only contributes $\approx$10\% of the total emission measure and thus has a small effect on the total flux. We discuss the hardness ratios in Appendix~\ref{app:HR}. In the ROSAT energy band of 0.1--2.4\,keV, convolution of the spectral model with the \texttt{cflux}  component yields a flux of $f_{\rm X} = 2.45^{+0.66}_{-0.58}\cdot10^{-14}\,\rm{erg\,cm^{-2}\,s^{-1}}$.

For both X-ray detections, we used the HD\hspace{0.05em}166620 \textit{Gaia} DR3 parallax of $\varpi=90.1234\,\rm{mas}$ \citep{Gaia_2023} to calculate the X-ray luminosity $L_{\rm X}$. The bolometric luminosity of ${\rm log}\,L_{\rm{bol}}\,\rm{(L_\sun)} = -0.45\pm0.02$ and a radius of $R_* = 0.80\pm0.02\,R_\sun$  \citep{Brewer_2016} were used to derive $R_{\rm X} = L_{\rm X}/L_{\rm bol}$ and the X-ray surface flux $F_{\!\rm X}$. The values for these X-ray activity diagnostics are presented in Table~\ref{tab:Xray_vals}, and $R_{\rm x}$ is overlaid on the $S_{\!{\rm HK}}$ time series in Fig.~\ref{fig:activity-timeseries}. The X-ray activity level of HD\,166620 has decreased by a factor of $2.5$ from its value close to the peak of the pre-MM chromospheric cycle to its  present MM-like state. 

\begin{table}[h!]
\centering
\caption{X-ray properties of HD\,166620.}
\label{tab:Xray_vals}
\resizebox{0.5\textwidth}{!}{
\begin{tabular}{c c c c}
\hline\hline
Observation & ${\rm log}\,L_{\rm X}$ & ${{\rm log}\,R_{\rm X}}$ & ${\rm log}\,F_{\!\rm X}$\\
 & $\rm{(erg\,s^{-1})}$ & & $\rm{(erg\,cm^{-2}\,s^{-1})}$\\
\hline\\[-10pt]
ROSAT/HRI & $26.96^{+0.12}_{-0.16}$ & $-6.17^{+0.12}_{-0.17}$ & $4.37^{+0.12}_{-0.17}$\\[2pt]
\textit{XMM-Newton} EPIC/pn & $26.56^{+0.10}_{-0.12}$ & $-6.58^{+0.10}_{-0.12}$ & $3.97^{+0.10}_{-0.12}$\\[2pt]
\hline\end{tabular}}
\end{table}

\begin{figure}
    \centering
    \includegraphics[width=0.45\textwidth]{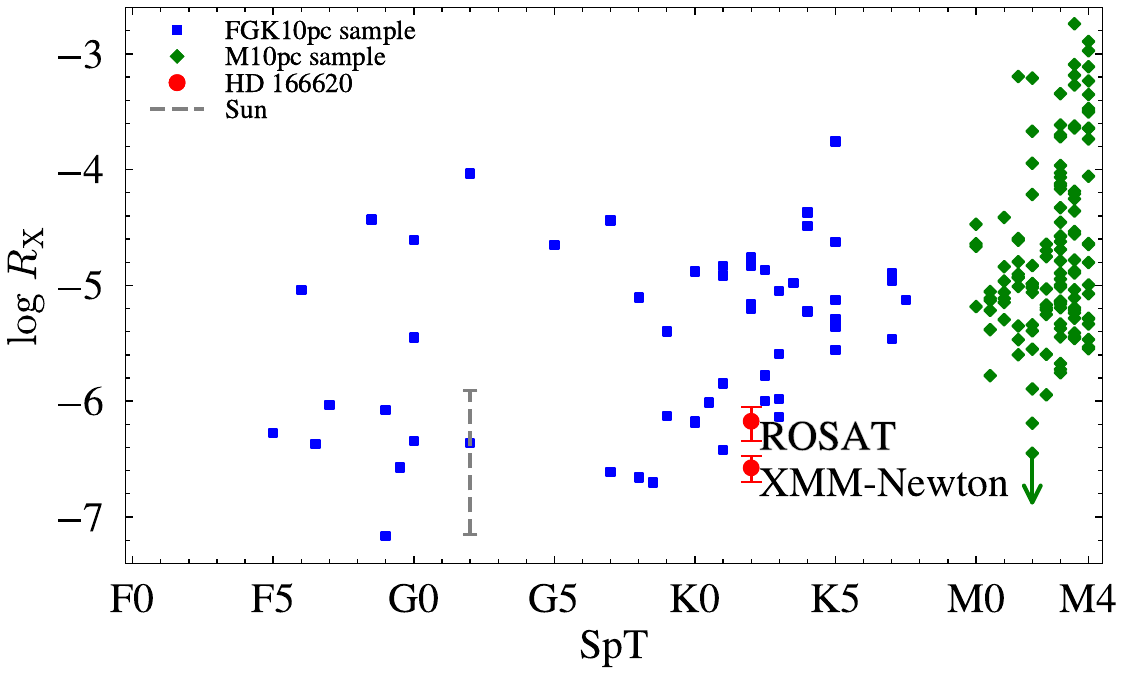}
    \caption{X-ray over bolometric luminosity for HD\,166620 in its two detections compared with the range of values observed throughout the solar cycle \citep[][]{Peres_2000} and for FGK (Bennedik et al., in prep.) and M0--M4 dwarfs \citep[][]{Caramazza_2023} within 10\,pc of the Sun.}
    \label{fig:Rx-plot}
\end{figure}

\subsection{Comparison sample}\label{subsect:sample}
For comparison, we refer to the M10pc sample described by \citet{Caramazza_2023}. Following a similar methodological approach, we compiled a sample of all FGK-type dwarfs within 10\,pc from the Sun (henceforth FGK\,10pc) based on the work of \citet{Reyle_2021} and the SpT therein (Bennedik et al., in prep.). The FGK\,10pc sample includes $64$ stars, $56$ of which have reliable \textit{Gaia} photometry. We collected X-ray data from ROSAT, eROSITA, and \textit{XMM-Newton} observations. To achieve this, we made use of the Second ROSAT All-Sky Survey Point Source Catalogue (2RXS) \citep{Boller_2016}, the Second ROSAT Source Catalogue of Pointed Observations (2RXP) \citep{Rosat_2000}, the eROSITA DR1 catalogue \citep{Merloni_2024}, and EPIC/pn detections in the 4XMM-DR14 catalogue \citep{Webb_2020}. In total, the FGK\,10pc sample contains X-ray data for $61$ stars, resulting in a completeness of >95\%. Within this sample, $4$ stars are marked as subgiants by \citet{Reyle_2021}, which we excluded from the analysis. 
The count rates listed in the X-ray catalogues were converted into a flux in the 0.1--2.4\,keV ROSAT energy band with proper conversion factors that depended on the instrumental setup (see Appendix~\ref{app:CFs}). With the parallaxes given by \citet{Reyle_2021}, we computed $L_{\rm X}$. 

As discussed in Appendix~\ref{app:Isochrones}, we used isochrones from the \textsc{parsec v2.0} database \citep{Nguyen_2022, Nguyen_2025} to estimate the bolometric luminosities and effective temperatures and the Stefan-Boltzmann law to determine the stellar radii. Finally, we computed $R_{\rm X}$ and $F_{\!\rm X}$ values for the FGK\,10pc sample.

\begin{figure}
    \centering
    \includegraphics[width=0.45\textwidth]{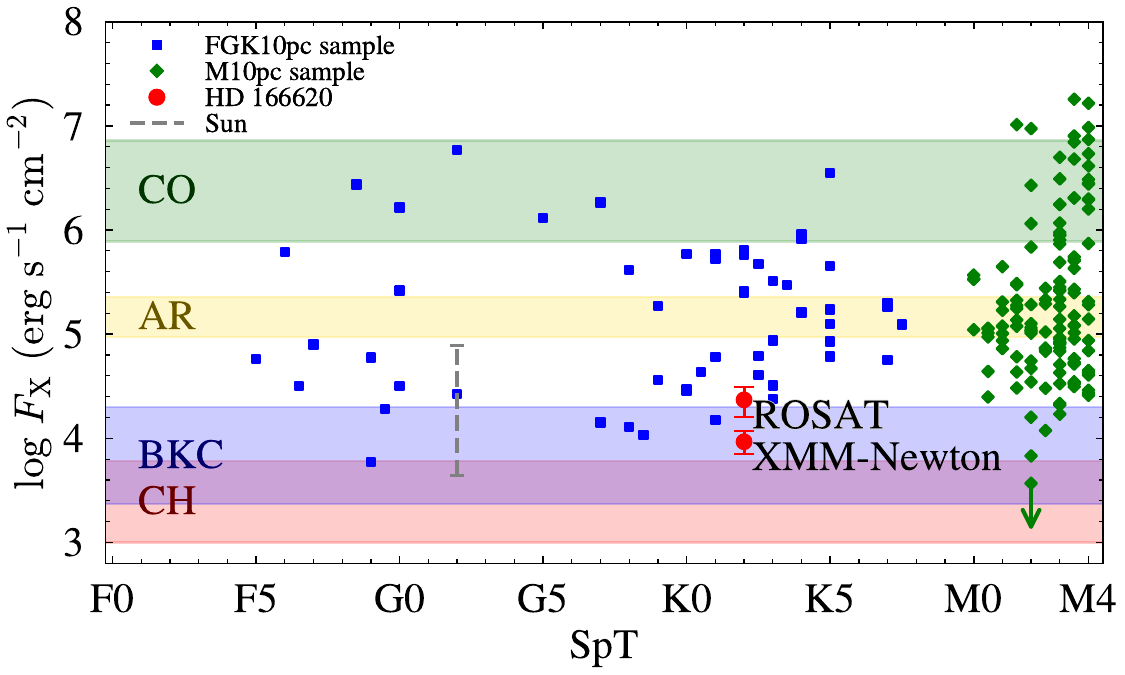}
    \caption{Same as Fig.~\ref{fig:Rx-plot}, but for X-ray surface fluxes $F_{\!\rm X}$. The typical ranges of $F_{\!\rm X}$ exhibited by different types of solar coronal structures \citep[from][]{Caramazza_2023} are depicted as coloured stripes.
    }
    \label{fig:Fx-plot}
\end{figure}

\section{Discussion}\label{sect:discussion}
Our systematic assessment of the stellar parameters and X-ray brightness of the FGK\,10pc sample allowed us to place the activity of HD\,166620 into context. In Fig.~\ref{fig:Rx-plot} we show the mean observed $R_{\rm X}$ values for each star in the FGK\,10pc sample and the M10pc\,sample from \citet{Caramazza_2023}. We display values for the X-ray observations of HD\,166620 and the solar ROSAT-band X-ray luminosities during its cycle extrema of $({L_{\rm{X,min}},L_{\rm{X,max}}}) = (0.27, 4.68)\cdot10^{27}\,\rm{erg\,s^{-1}}$ by \citet{Peres_2000}.

Fig.~\ref{fig:Rx-plot} shows that the X-ray emission of FGK dwarfs spans at least three orders of magnitude. This was known from previous systematic studies such as \citet{Schmitt_2004}, who used a smaller sample in the ROSAT era, and \citet{Zhu_2025}, who presented a compilation of X-ray data for a larger but less complete sample of GKM stars. The ROSAT detection places HD\,166620 at the lower end of the observed $R_{\rm X}$ range of K dwarfs, even though this measurement was taken close to the maximum of its chromospheric cycle. After the onset of the MM-like state, the $R_{\rm X}$ value of HD\,166620 has dropped below that seen in all other K dwarfs within the 10\,pc sample. The rising slope of the lower envelope in Fig.~\ref{fig:Rx-plot} is likely due to the longer spin-down timescales of later-type stars, which delay their development to low activity levels \citep{Preibisch_2005, Johnstone_2021}. Thus, the position of HD\,166620 at the low activity end is in line with its slow rotation \citep[45\,d,][]{Luhn_2022} and old age \citep[$12.4\,\rm{Gyr}$,][]{Brewer_2016}.

The X-ray surface flux, $F_{\!\rm X}$, is a parameter that offers the possibility to directly compare stellar and solar activity levels since it is independent of the size of the emitting region. Extending Fig.~3 of \citet{Caramazza_2023}, we show in Fig.~\ref{fig:Fx-plot} the $F_{\!\rm X}$ of HD\,166620, the stars in the FGK\,10pc sample, and the M10pc sample together with the typical values of different types of solar coronal magnetic structures. The latter, divided into solar coronal holes (CH), background corona (BKC), active regions (AR), and cores of active regions (CO), were derived from data from the {\it Yohkoh} satellite, originally as part of the {\it Sun as an X-ray star} project \citep{Orlando_2001, Orlando_2004}, and their $F_{\!\rm X}$ values were presented by \citet{Caramazza_2023}.

While at the time of the ROSAT observation, the $F_{\!\rm X}$ of HD\,166620 was at the upper edge of the values exhibited by solar BKC, during its current MM-like state, the HD\,166620 corona is fully consistent with the emission levels of solar BKC and just slightly above the upper boundary for CH. The solar BKC consists of quiet magnetic loops that trap relatively low-temperature ($\approx 1$\,MK) plasma \citep{Orlando_2001, Reale_2014}. The X-ray plasma temperature of $0.1$\,keV derived from the {\it XMM-Newton} spectrum of HD\,166620 is coincident with this value, which further strengthens the hypothesis that during its current MM state, the star resembles the quiet Sun. The surface-averaged X-ray flux of the Sun itself \citep{Peres_2000} is in the range of the BKC throughout the low-activity part of its $11$-year cycle. Full-disk images from solar instruments such as {\it Yohkoh} or {\it Hinode} show the absence of high-activity magnetic structures such as CO and flares during solar minima \citep{Adithya_2021}.

When HD\,166620 is considered as a template for the Sun, our results provide evidence that the solar activity level during the historical MM was not reduced far below the level seen during its modern-time cycle minima. This agrees with direct coronal observations of solar eclipses in the historical MM, which show a weak and unstructured corona. If the Sun had been covered with large magnetic structures, it is likely that these would have been seen in reports and paintings at the time (see \citet{Hayakawa_2021, Usoskin_2015} for such historical observations). 

The 10\,pc sample shows that HD\,166620 during its MM state and the Sun during its cycle minimum appear to define a natural lower boundary of the surface-averaged coronal X-ray flux; with the exception of very rare coronal-hole stars  \citep{Caramazza_2023}. This flux is provided by the small quiet magnetic loops of the solar(-like) BKC on both stars. Despite its later SpT, HD\,166620 is a close analogue of the Sun in terms of its large-scale magnetic field as well. Specifically, both stars are located at the critical Rossby number that identifies stars with weak dipolar field due to low dynamo efficiency  \citep{Metcalfe_2025}. In simulations of the Babcock-Leighton dynamo, slowly rotating stars produce magnetic cycles with long-term variations of the strength of the dipole, and the low dynamo number of these stars makes their recovery from phases of a weak cycle slow, which in turn leads to grand minima \citep{Vashishth_2023}. Other stars with a similar BKC-like X-ray surface flux as HD\,166620 and the Sun can therefore be expected to undergo MM states. These stars can be identified as the objects in the weakened magnetic braking regime \citep{van_Saders_2016,Metcalfe_2025}. 

\section{Summary and conclusions}\label{sect:summary}

The K2 dwarf HD\,166620 is the only star that is unambiguously known to be in an MM-like activity state. Our \textit{XMM-Newton} observation is the first X-ray detection since the onset of its diminished activity in approximately 2004. In its current MM-like state, the X-ray emission of HD\,166620 has decreased by a factor of $2.5$ compared to its pre-MM state, and it lies at ${{\rm log}\,L_{\rm X}\,\rm{(erg\,s^{-1})}=26.56^{+0.10}_{-0.12}}$ and ${R_{\rm X} = -6.58^{+0.10}_{-0.12}}$. This is slightly below the lower envelope defined by the other K dwarfs within 10\,pc. The X-ray surface flux of HD\,166620, ${{\rm log}\,F_{\!\rm X}\,\rm{(erg\,cm^{-2}\,s^{-1})} = 3.97^{+0.10}_{-0.12}}$, is consistent with the emission levels typically seen in the solar background corona, which dominates the minimum of the solar cycle. Together with the $S_{\!{\rm HK}}$ time series of HD\,166620, this suggests that during the MM of the Sun, the magnetic activity was not reduced significantly below the levels seen during its current cycle minima.

\begin{acknowledgements} 
We thank the anonymous referee for their feedback and suggestions.
MB and MC acknowledge support by the Bundesministerium für Wirtschaft und Energie through the Deutsches Zentrum für Luft- und Raumfahrt (DLR) under grant numbers FKZ 50 OR 2505 and FKZ 50 OR 2316.
This research has made use of data obtained from the 4XMM {\it XMM-Newton} serendipitous source catalogue compiled by the $10$ institutes of the {\it XMM-Newton} Survey Science Centre selected by ESA, and of archival data of the ROSAT space mission.
This work is also based on data from eROSITA, the soft X-ray instrument aboard SRG, a joint Russian-German science mission supported by the Russian Space Agency (Roskosmos), in the interests of the Russian Academy of Sciences represented by its Space Research Institute (IKI), and the DLR. The SRG spacecraft was built by Lavochkin Association (NPOL) and its subcontractors, and is operated by NPOL with support from the Max Planck Institute for Extraterrestrial Physics (MPE). The development and construction of the eROSITA X-ray instrument was led by MPE, with contributions from the Dr. Karl Remeis Observatory Bamberg \& ECAP (FAU Erlangen-Nuernberg), the University of Hamburg Observatory, the Leibniz Institute for Astrophysics Potsdam (AIP), and the Institute for Astronomy and Astrophysics of the University of Tübingen, with the support of DLR and the Max Planck Society. The Argelander Institute for Astronomy of the University of Bonn and the Ludwig Maximilians Universität Munich also participated in the science preparation for eROSITA.
This research has made use of data and software provided by the High Energy Astrophysics Science Archive Research Center (HEASARC), which is a service of the Astrophysics Science Division at NASA/GSFC.
This work has made use of data from the ESA mission {\it Gaia} (\url{https://www.cosmos.esa.int/gaia}), processed by the {\it Gaia} Data Processing and Analysis Consortium (DPAC, \url{https://www.cosmos.esa.int/web/gaia/dpac/consortium}). Funding for the DPAC has been provided by national institutions, in particular the institutions participating in the {\it Gaia} Multilateral Agreement.
      
\end{acknowledgements}

   \bibliographystyle{aa}
   \bibliography{Maunder_biblio}

\begin{appendix} 

\section{$S_{\rm HK}$ from Automated Planet Finder}\label{app:SHK-APF}

\begin{figure}[h!]
    \centering
    \includegraphics[width=0.48\textwidth]{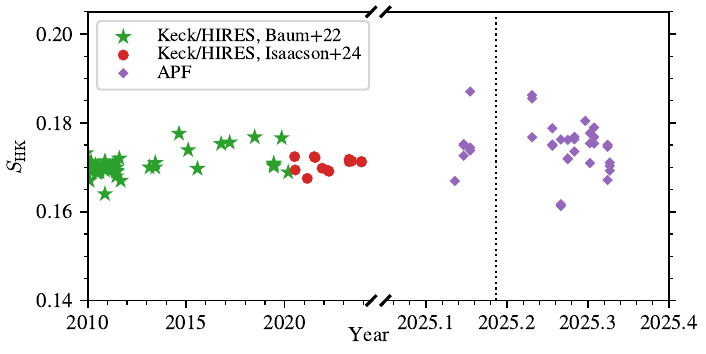}
    \caption{$S_{\!{\rm HK}}$ time series for HD\,166620 showing the Keck/HIRES data in comparison with the newly obtained APF data. The dotted vertical line marks the time of the \textit{XMM-Newton} observation.
    }
\label{fig:SHK-APF}
\end{figure}

We obtained $S_{\!{\rm HK}}$ measurements of HD~166620 taken with the APF \citep{Vogt_2014} around the time of our {\it XMM-Newton} observation. In Fig.~\ref{fig:SHK-APF}, we compare these data with the most recent $S_{\!{\rm HK}}$ measurements from Keck/HIRES. For both the APF and Keck/HIRES, the calibration is performed using the same method, which involves insertion of a cell of gaseous iodine in the light beam \citep[see e.g.][]{Toledo_Padron_2019}. However, the APF data shows a higher scatter due to the lower sensitivity of the APF detector near the Ca\,II~H\&K lines and the smaller aperture of the instrument. Throughout the APF monitoring near the time of our \textit{XMM-Newton} observation in early 2025, HD\,166620 has stayed at a constantly low $S_{\!{\rm HK}}$ activity level. The mean $S_{\!{\rm HK}}$ value from the two-month span of APF monitoring is within $1\,\sigma$ of the most recent  Keck/HIRES data.

\section{Hardness ratios of HD\,166620}\label{app:HR}

With only 87 net source counts, spectral fitting is at its limits. In order to verify the plausibility of the spectral model for the \textit{XMM-Newton} EPIC/pn observation of HD\,166620 in Sect.~\ref{subsect:data-Xray}, we calculated hardness ratios for the soft (S), medium (M) and hard (H) bands which span the energy ranges 0.2--0.5\,keV, 0.5--1.0\,keV, and 1.0--2.0\,keV, respectively. The hardness ratios are defined as ${HR_1 = \frac{R_M - R_S}{R_M + R_S}}$ and ${HR_2 = \frac{R_H - R_M}{R_H + R_M}}$, where $R_i$ is the count rate for energy band $i$.

The observed hardness ratios of HD\,166620 are ${HR_1 = -0.44 \pm 0.13}$ and ${HR_2 = -1^{+0.17}_{-0.00}}$. Using XSPEC, we created a simulated APEC spectrum with the EPIC/pn instrumental response files of the observation for the dominating component from the spectral fit at a temperature of ${kT = 0.109^{+0.019}_{-0.015} \,\rm{keV}}$ (see Sect.~\ref{subsect:data-Xray}). For the simulated spectrum, we obtained  the count rate for each of the aforementioned energy bands and we computed  the ``expected" hardness ratios for a star with that mean coronal temperature. The resulting values of ${HR_1^{\rm{sim}} = -0.55}$ and ${HR_2^{\rm{sim}} = -0.99}$ are consistent with the observed values, showing the validity of the spectral model.

\section{Rate-to-flux conversion factors}\label{app:CFs}

Catalogued X-ray count rates in energy band $i$, $\si{CR}_{\rm i}$, can be converted into a flux in energy band $j$, $f_{\rm j}$, if the corresponding conversion factor ($\si{CF_{ij}}$) is known. The $\si{CF_{ij}}$ is computed as

\begin{equation}
    \si{CF_{ij}} = \frac{f_j}{\si{CR}_i}.
\end{equation} 

The CF depends on the characteristics of the instrument and the spectral shape of the observed emission. In the X-ray catalogues used in this article, the source count rates have been corrected for PSF-loss and vignetting effects and are thus handled as on-axis count rates with an infinite extraction radius.

We simulate the typical X-ray spectrum of an FGK dwarf in XSPEC, using the best-fit model determined from an analysis of the average eRASS spectrum of all FGK dwarfs within 10\,parsec \citep{Zheng_2025}. This model is a three-temperature variable abundance (3T-VAPEC) model. Using the \texttt{flux} routine in XSPEC, we obtain the flux in the 0.1--2.4\,keV ROSAT energy band from the model. Then we use the \texttt{fakeit} routine to convolve the simulated spectrum with the specific response matrix of each instrument to compute an expected count rate for the energy band used in the catalogue.

In the following, we briefly discuss considerations for each instrument used. All calculated CFs are presented in Table~\ref{tab:CFs}.

\begin{table}[h!]
\centering
\caption{Instrument-specific CFs to convert count rates for FGK dwarfs into 0.1--2.4\,keV ROSAT band fluxes.}
\label{tab:CFs}
\resizebox{0.5\textwidth}{!}{
\begin{tabular}{l c c c}
\hline\hline
Instrument & obs. energy band & $\si{CF_{FGK}}$\\
& (keV) & {($10^{-12}\,\rm{erg\,cm^{-2}\,cts^{-1}}$)}\\
\hline
ROSAT PSPC-C & 0.1--2.4 & 5.77\\
ROSAT PSPC-B (before 1991 Oct 14) & 0.1--2.4 & 5.87\\
ROSAT PSPC-B (after 1991 Oct 14) & 0.1--2.4 & 5.92\\
eROSITA & 0.2--2.3 & 1.06\\
\textit{XMM-Newton} EPIC/pn; thin Filter & 0.2--12.0 & 1.60\\
\textit{XMM-Newton} EPIC/pn; medium Filter & 0.2--12.0 & 1.71\\
\textit{XMM-Newton} EPIC/pn; thick Filter & 0.2--12.0 & 2.35\\
\hline
\end{tabular}
}
\end{table}

\subsection*{ROSAT}
For 2RXS observations the instrument PSPC-C was used. We use the response file \texttt{pspcc\_gain1\_256.rsp} from HEASARC\footnote{\url{https://heasarc.gsfc.nasa.gov/docs/rosat/pspc\_matrices.html}}, which already contains the on-axis effective area.

In our sample, all 2RXP observations use the PSPC-B detector. For observations before the destruction of the PSPC-C and the corresponding change in detector gain of PSPC-B on 1991-10-14, the appropriate file is \texttt{pspcb\_gain1\_256.rsp}. For later observations, we use \texttt{pspcb\_gain2\_256.rsp}.

\subsection*{eROSITA}
As part of the first eROSITA data release \citep{Merloni_2024}, the calibration products were made public. For the catalogue count rates, the appropriate files are provided as \texttt{onaxis\_tm0\_rmf\_2023-01-17.fits} and \texttt{onaxis\_tm0\_arf\_filter\_2023-01-17.fits}.

\subsection*{XMM-Newton}
In this article, we only consider count rates from EPIC/pn. ESA provides `canned' response matrices\footnote{\url{https://www.cosmos.esa.int/web/xmm-newton/epic-response-files}}. Catalogue values are provided for ``single + double'' imaging patterns, and the appropriate file is \texttt{epn\_e4\_ff20\_sdY9\_v22.0.rmf}. 

The on-axis ancillary files are generated with the \texttt{arfgen} command in SAS. To mitigate PSF-losses, we choose a large extraction region of 5\,arcmin. The EPIC/pn observations of our sample were carried out using the ``thick" or the ``medium" filter. The different transmissions of the EPIC/pn filters are taken into account with individual ancillary files. For completeness, we provide the CF for the ``thin" filter, too.

\subsection*{Cross-calibration}
\begin{figure}[h!]
    \centering
    \includegraphics[width=0.4\textwidth]{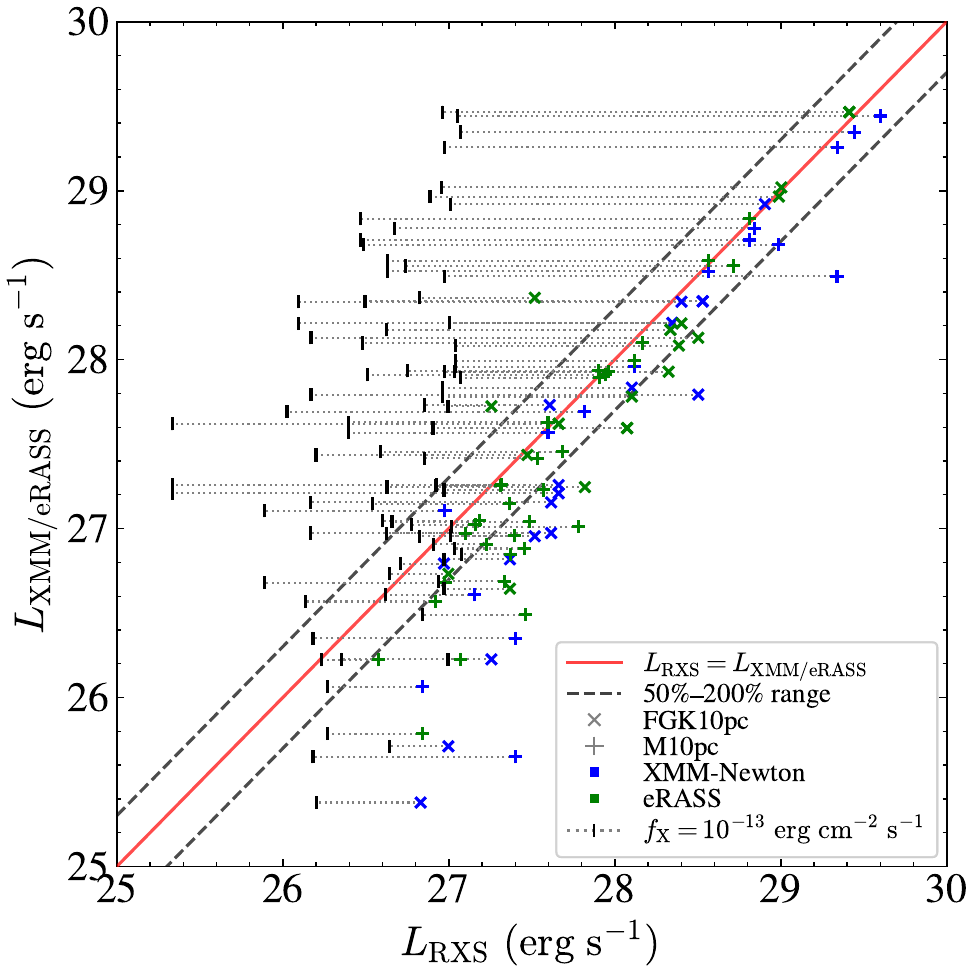}
    \caption{Comparison of X-ray luminosities detected with 2RXS, eROSITA, and \textit{XMM-Newton} in the ROSAT energy band after applying the respective CF shown in Table~\ref{tab:CFs}. For the M10pc sample, we use the CFs from \citet{Caramazza_2023}. Black bars represent the sensitivity limit of 2RXS at the individual distances of the stars.}
\label{fig:CrossCalibration}
\end{figure}

Stars in our sample that were observed by multiple instruments allow us to verify the CFs by comparing the X-ray luminosities derived from the different instruments, as presented in Fig.~\ref{fig:CrossCalibration}. 
We notice a systematic trend where objects appear brighter in ROSAT observations compared to eROSITA or \textit{XMM-Newton}.

This trend is most significant for fainter objects (${{\rm log}\,L_{\rm X,2RXS}\,{\rm (erg\,s^{-1})} \lesssim 27.5}$) where the difference can be explained by an observational bias due to the lower sensitivity of ROSAT, combined with intrinsic variability of the stars. Namely, X-ray faint stars (lower left of the diagram) were detectable in 2RXS only if they were brighter than  during their observation with  {\it XMM-Newton} or eRASS.
This is illustrated in Fig.~\ref{fig:CrossCalibration} with black vertical dashes, which represent the individual approximate 2RXS X-ray luminosity limits, derived from the RASS flux limit of ${f_{\rm X} \approx 10^{-13}\,{\rm erg\, cm^{-2}\, s^{-1}}}$ \citep{Boller_2016}. For the faintest stars this limit is above their $L_{\rm X}$ value measured with the more sensitive instruments, \textit{XMM-Newton} or eROSITA. Similar observations were made on other samples by,  e.g., \citet{Magaudda_2022, Freund_2018, Zhu_2025}.
 
A systematic offset is also seen for the brighter regime (${{\rm log}\,L_{\rm X,2RXS}\,{\rm (erg\,s^{-1})} \gtrsim 27.5}$), albeit at a smaller scale. A possible origin for this trend might be calibration differences not fully captured in the instrument response files. Moreover, the spectral model of the average FGK dwarf \citep{Zheng_2025} used to compute the CFs was obtained from a fit to the eRASS spectrum in the 0.35--2.0\,keV energy band and had to be extrapolated to cover the full ROSAT energy band which may lead to an underpredicted soft emission. The spectral 3T-VAPEC model of the average FGK dwarf has an emission-weighted mean temperature of ${\overline{kT}=0.417\,\rm{keV}}$. For stars with cooler coronae, this could lead to an underestimated flux in the low  energy range. In fact, the dominating spectral component of HD\,166620 is at a temperature of ${kT = 0.109^{+0.019}_{-0.015} \,\rm{keV}}$ and its effective conversion factor from the count rates of source detection to the flux in the ROSAT-band is ${7.21\cdot10^{-12}\,\rm{erg\,cm^{-2}\,cts^{-1}}}$, which is a factor 4.2 higher than the CF determined for a ``medium'' filter \textit{XMM-Newton} observation. While HD\,166620 is an extreme example, it shows that for other FGK dwarfs with very soft X-ray emission, the CFs for \textit{XMM-Newton} and eROSITA may be underestimated. In that case, the true X-ray luminosities of such stars observed by \textit{XMM-Newton} and eROSITA would be higher, putting them closer to the values observed with ROSAT.

\section{Isochrone fitting}\label{app:Isochrones}

We obtain fundamental stellar parameters for our sample from fitting isochrones on the colour-magnitude diagram. The isochrones are adopted from the \textsc{parsec v2.0} database \citep{Nguyen_2022, Nguyen_2025}, without rotation. To perform the fit, we use \textit{Gaia} colours where available, and Johnson B-V-colours otherwise. A grid of isochrones is created with ages spanning from 0.5 to 9.0\,Gyr with a step-size of 0.5\,Gyr, and metallicities ([M/H]) in the range from -1.5 to 0.5\,dex with a step-size of 0.1\,dex. We use metallicities from \citet{Soubiran_2022} where available and else from \citet{Gaspar_2016, Sousa_2008} to constrain the isochrones, while leaving the age free to be fit. We assume that [Fe/H]\,$\approx$\,[M/H] as the sample is limited to the nearby solar neighbourhood, which is part of the thin disk and thus not significantly alpha-enhanced \citep{Bensby_2014, Buder_2019}.

For an isochrone of matching metallicity, the best-fit value is determined by the minimum distance between the observed data and the points of the isochrone in the colour-magnitude-space. In case there is no exact match with the observed metallicity due to the limited resolution of the grid, we linearly interpolate between the best-fit values of the two closest metallicities. As a result, we obtain fundamental parameters such as the bolometric luminosity and effective temperature of the star. 

\begin{figure}[h!]
    \centering
    \includegraphics[width=0.385\textwidth]{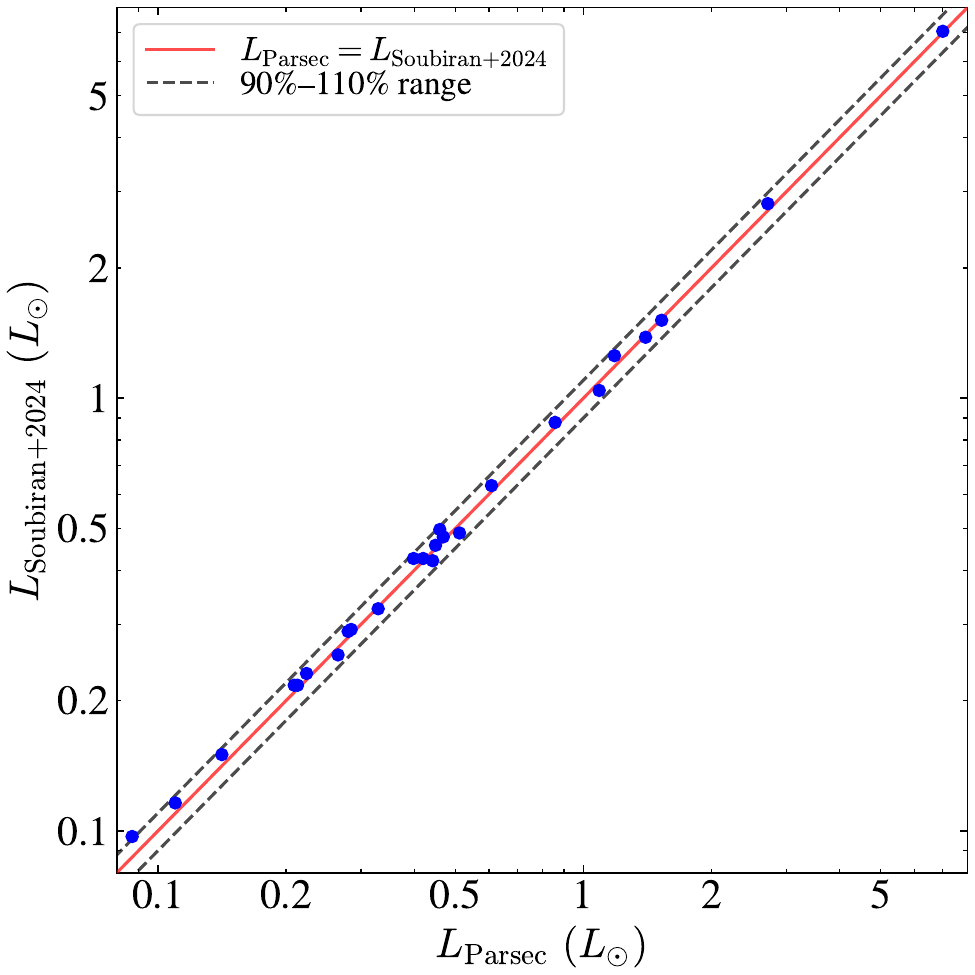}
    \caption{Bolometric luminosities for our sample obtained with \textsc{parsec v2.0} compared with those published by \citet{Soubiran_2024}. The dashed lines mark a deviation of 10\%.
    }
    \label{fig:Comparison-Lbol-Soubiran}
\end{figure}

From the FGK\,10pc sample, 25 dwarfs are also part of the \textit{Gaia} FGK benchmark stars \citep{Soubiran_2024}. A comparison of our values for the bolometric luminosity with the values from \citet{Soubiran_2024} shows an agreement well within 10\%, as depicted in Fig.~\ref{fig:Comparison-Lbol-Soubiran}.

\end{appendix} 

\end{document}